\documentclass[sigconf]{acmart}
\AtBeginDocument{%
  }

\copyrightyear{2026}
\acmYear{2026}
\setcopyright{cc}
\setcctype{by}
\acmConference[ICPC '26]{34th IEEE/ACM International Conference on Program Comprehension}{April 12--13, 2026}{Rio de Janeiro, Brazil}
\acmBooktitle{34th IEEE/ACM International Conference on Program Comprehension (ICPC '26), April 12--13, 2026, Rio de Janeiro, Brazil}
\acmPrice{}
\acmDOI{10.1145/3794763.3794803}
\acmISBN{979-8-4007-2482-4/2026/04}

\usepackage{booktabs}
\usepackage{multirow}
\usepackage{tcolorbox}
\usepackage{booktabs}
\usepackage{pifont}
\usepackage{makecell}
\usepackage{caption}
\usepackage{tabularx}
\usepackage{amsmath}
\usepackage{booktabs}   
\usepackage{tabularx}   
\usepackage{makecell}   
\usepackage{pifont}     
\usepackage{array}      
\usepackage{caption}    

\usepackage{booktabs} 
\usepackage{multirow} 

\newcommand{\cmark}{\checkmark}
\newcommand{\xmark}{\ding{55}}

\begin{document}



\title[Aligning Large Language Models for SQL Comment Generation with Direct Preference Optimization]{SQL-Commenter: Aligning Large Language Models for SQL Comment Generation with Direct Preference Optimization}


\author{Lei Yu}
\affiliation{%
  \institution{Institute of Software, Chinese Academy of Sciences, University of Chinese Academy of Sciences}
  \city{Beijing}
  \country{China}}
\email{yulei2022@iscas.ac.cn}

\author{Peng Wang}
\affiliation{%
  \institution{Institute of Software, Chinese Academy of Sciences, University of Chinese Academy of Sciences}
  \city{Beijing}
  \country{China}}
\email{wangpeng232@mails.ucas.ac.cn}

\author{Jingyuan Zhang}
\affiliation{%
  \institution{Institute of Software, Chinese Academy of Sciences, University of Chinese Academy of Sciences}
  \city{Beijing}
  \country{China}}
\email{zhangjingyuan2023@iscas.ac.cn}

\author{Xin Wang}
\affiliation{%
  \institution{Institute of Software, Chinese Academy of Sciences, University of Chinese Academy of Sciences}
  \city{Beijing}
  \country{China}}
\email{wangxin@iscas.ac.cn}

\author{Jia Xu}
\affiliation{%
  \institution{Institute of Software, Chinese Academy of Sciences, University of Chinese Academy of Sciences}
  \city{Beijing}
  \country{China}}
\email{xujia23@mails.ucas.ac.cn}

\author{Li Yang}
\authornote{Corresponding authors}
\affiliation{%
  \institution{Institute of Software, Chinese Academy of Sciences}
  \city{Beijing}
  \country{China}}
\email{yangli2017@iscas.ac.cn}

\author{Changzhi Deng}
\affiliation{%
  \institution{Institute of Software, Chinese Academy of Sciences}
  \city{Beijing}
  \country{China}}
\email{dengchangzhi@iscas.ac.cn}

\author{Jiajia Ma}
\affiliation{%
  \institution{Institute of Software, Chinese Academy of Sciences}
  \city{Beijing}
  \country{China}}
\email{majiajia@iscas.ac.cn}

\author{Fengjun Zhang}
\authornotemark[1]
\affiliation{%
  \institution{Institute of Software, Chinese Academy of Sciences}
  \city{Beijing}
  \country{China}}
\email{fengjun@iscas.ac.cn}

\renewcommand{\shortauthors}{Yu et al.}

\begin{abstract}
    SQL query comprehension is a significant challenge in database and data analysis environments due to complex syntax, diverse join types, and deep nesting. Despite its critical role in backend development and data science, many queries, particularly within legacy systems, often lack adequate comments, which severely hinders code readability, maintainability, and knowledge transfer. Existing approaches to automated SQL comment generation face two main challenges: limited training datasets that inadequately represent real-world analytical queries involving multi-table joins, window functions, and complex aggregations, and an insufficient understanding of SQL-specific logical semantics and schema-related context by Large Language Models (LLMs), even after standard training. Our empirical analysis shows that even after continual pre-training and supervised fine-tuning, LLMs struggle to precisely understand complex SQL semantics, leading to inaccurate or incomplete comments. To address these challenges, we propose SQL-Commenter, an advanced comment generation method based on LLaMA-3.1-8B. First, we construct a comprehensive dataset containing longer, more complex SQL queries with expert-verified, detailed comments. Second, we perform continual pre-training using a large-scale SQL corpus to enhance the LLM’s understanding of SQL syntax and semantics. Then, we conduct supervised fine-tuning with our high-quality dataset. Finally, we introduce Direct Preference Optimization (DPO), which leverages human feedback to significantly improve comment quality. SQL-Commenter utilizes a preference-based loss function that encourages the LLM to increase the probability of preferred outputs while decreasing the probability of non-preferred outputs, thereby enhancing both fine-grained semantic learning, such as distinguishing between different join types, and context-dependent quality assessment based on business logic. We evaluate SQL-Commenter on the authoritative Spider and Bird benchmarks, where it significantly outperforms state-of-the-art baselines. On average, across these datasets, our method surpasses the strongest baseline (Qwen3-14B) by 9.29, 4.99, and 13.23 percentage points on BLEU-4, METEOR, and ROUGE-L, respectively. Moreover, human evaluation demonstrates the superior quality of comments generated by SQL-Commenter in terms of correctness, completeness, and naturalness.
\end{abstract}

\ccsdesc[500]{Software and its engineering}
\ccsdesc[300]{Computing methodologies~Natural language processing}

\keywords{SQL, Large Language Models, Direct Preference Optimization, Comment Generation}

\received{23 October 2025}
\received[accepted]{26 January 2026}

\maketitle

\section{Introduction}

SQL (Structured Query Language) remains one of the most widely adopted programming languages, utilized by over half (51.52\%) of professional developers \cite{hong2025next}. However, its syntactic complexity, manifested through intricate join operations, deeply nested subqueries, and complex business logic, creates substantial comprehension challenges, even for experienced practitioners \cite{QuoraSQLDifficulty}. Exacerbating this challenge is the finding that only about a third (35.29\%) of practitioners report receiving systematic training \cite{StackOverflowSurvey2023}. This problem is particularly acute in large-scale legacy systems, where SQL queries often lack adequate comments, severely hindering code readability, maintainability, and knowledge transfer. The consequences of misinterpreting SQL queries are significant, potentially leading to operational inefficiencies \cite{yan2017understanding}, security vulnerabilities \cite{thomas2009automated}, and data integrity issues \cite{muse2020prevalence}. The urgency and scale of this comprehension barrier are starkly illustrated on platforms like Stack Overflow; as of October 19, 2025, a search for the "sql" tag reveals over 674,654 related questions, highlighting a persistent and widespread need for assistance \cite{StackOverflowSQLQuestions}. Translating SQL queries into natural language (SQL-to-Text) offers a practical solution to these challenges \cite{ma2021relation}. For technical professionals, it provides detailed explanations to verify logic and mitigate risks, while for non-technical users, it enhances the accessibility and usability of databases. Despite this clear need, and unlike general-purpose programming languages (GPPLs) where automated comment generation is well-studied \cite{geng2024large, li2022automating, lu2023llama, lu2025deepcrceval}, domain-specific languages (DSLs) such as SQL remain largely unexplored in this context, presenting a significant research gap.

Recent benchmarks for Large Language Models (LLMs) have explicitly evaluated the SQL-to-Text task, but their framing reveals a fundamental limitation. For instance, a comprehensive study by Zhang et al. \cite{zhang2024benchmarking} assesses various LLMs (Codellama \cite{roziere2023code}, InternLM \cite{team2023internlm}, InternLM2 \cite{cai2024internlm2}, Llama2 \cite{touvron2023llama}) on their ability to transform a SQL query back into its "original natural language question" \cite{shu2021logic}. A critical observation is that these models, whether general-purpose (Llama2, InternLM) or code-specialized (Codellama), are used off-the-shelf without any specialized training or fine-tuning for the SQL-to-Text task. Their evaluation, based on metrics like ROUGE \cite{lin2004rouge} and BertScore \cite{zhang2019bertscore}, focuses on how closely the generated text matches the user's initial query. This "SQL-to-Question" task demonstrates an LLM's semantic understanding but fundamentally fails to address a developer's needs. The output is a high-level summary of the query's goal, not a technical explanation of its execution, omitting crucial details about join strategies, subquery mechanics, or complex filtering logic. On the other hand, while graph-based models (Graph2Seq, GNNs, HeSQLNet) attempted to better capture SQL's structure by parsing it into an Abstract Syntax Tree (AST) \cite{xu2018sql, ma2021relation, zhang2025hesqlnet}, the 'text' they generate is similarly confined to high-level questions or descriptions, sharing the same fundamental limitation. Furthermore, these methods introduce their own complexities, requiring a rigid, intermediate graph representation that can be brittle and may fail to capture all semantic nuances, leading to structural ambiguities. For these reasons, our approach bypasses complex graph-encoding and instead enhances the inherent code understanding of a LLM through targeted post-training, enabling it to generate true technical explanations, not just questions.


LLMs have garnered attention for their powerful natural language understanding and generation capabilities, demonstrating remarkable performance in various complex tasks, including code comprehension and generation \cite{li2023can, liu2023your, vaithilingam2022expectation, weyssow2025exploring, lu2023llama, cheng2025auvana, shen2024dependency, zan2024swe}, smart contract auditing and generation \cite{yu2024smart, yu2025smart, yu2023pscvfinder, yu2025sael, yuan2025mos, yu2025towards, yuan2025leveraging}. However, they often encounter significant difficulties in handling SQL-specific concepts and schema-related context, a challenge starkly illustrated by the example in Figure \ref{tab: motivation}. When presented with a complex query to find a card from the artist who created the most promotional items, a general-purpose model like LLaMA-3.1-8B-Instruct completely misunderstands its logic, incorrectly inferring that the goal is to find the card with the "most ruling information." While a specialized model after continual pre-training and supervised fine-tuning (CPT+SFT) shows improvement by correctly identifying the subquery's purpose, it still produces deficient comments. As shown in the figure, it fails to grasp the holistic intent, making a subtle but critical semantic error by claiming the query then selects the card with the "most ruling information from this artist," which implies a non-existent aggregation. These persistent errors, progressing from overt factual mistakes in general models to subtle semantic misunderstandings in fine-tuned ones, demonstrate that traditional training approaches (CPT+SFT) are insufficient for developing models with a deep and accurate understanding of complex SQL semantics.

To address these challenges, we propose SQL-Commenter, based on the LLaMA-3.1-8B model. This approach combines continual pre-training (CPT), supervised fine-tuning (SFT), and direct preference optimization (DPO). By utilizing a large-scale SQL corpus for domain-specific pre-training, we enable the LLM to better understand the syntax and semantics of SQL. To address the limitations of existing datasets, which often lack real-world analytical complexity, we construct a comprehensive dataset containing longer, more complex SQL queries with expert-verified, detailed comments, drawing from benchmarks like Spider and Bird. The SFT process using this high-quality dataset enhances the LLM's capability to annotate complex queries involving multi-table joins and window functions. Finally, to address the subtle issues in comment quality that persist even after CPT and SFT, we introduce DPO. DPO improves comment quality through two key aspects: (1) Fine-grained Semantic Learning, capturing query-specific details such as distinguishing between different join types or correctly interpreting aggregate functions with HAVING clauses; (2) Context-dependent Quality Assessment, learning to generate comments that reflect the underlying business logic, not just the technical operations. By constructing a dataset of preferred and non-preferred comment pairs, DPO enables the LLM to generate higher-quality comments that better align with human expert expectations.

Our data construction involved a two-stage process. First, for the SFT dataset, we prompted DeepSeek-V3.1 (non-thinking mode) with comprehensive details from Text-to-SQL benchmarks (including the query, question, and schema) to generate initial explanatory comments. These were then meticulously refined by a team of experienced data analysts and DBAs. For the DPO dataset, these expert-refined comments became the 'chosen' responses. The corresponding 'rejected' responses were generated by guiding DeepSeek-V3.1 (non-thinking mode) through a multi-strategy negative sampling pipeline to intentionally introduce flaws, such as technical errors, superficiality, or incompleteness, thereby creating a distinct quality gap for preference learning.



Experimental results on the Spider and Bird benchmarks confirm SQL-Commenter’s superior performance. On the Spider dev set, it achieved state-of-the-art scores of 36.95\% BLEU-4, 58.37\% METEOR, and 57.17\% ROUGE-L, representing improvements of 9.02, 4.80, and 13.41 percentage points over the strongest baseline, Qwen3-14B. This strong performance is mirrored on the Spider test set, where it scored 36.37\% BLEU-4, 57.76\% METEOR, and 56.48\% ROUGE-L, surpassing the same baseline by 9.48, 5.80, and 12.03 percentage points, respectively. On the more challenging Bird dev set, SQL-Commenter continues to lead, achieving 35.09\% BLEU-4, 55.91\% METEOR, and 56.74\% ROUGE-L, which are 9.36, 4.37, and 14.24 percentage points higher than the Qwen3-14B baseline. Moreover, human evaluation demonstrates the superior quality of comments generated by SQL-Commenter in terms of correctness, completeness, and naturalness, significantly outperforming all baselines.

The main contributions of this paper are as follows:
\begin{itemize}
\item To the best of our knowledge, we are the first to introduce Direct Preference Optimization in the SQL comment generation task.
\item We publicly released comprehensive datasets for training and evaluating SQL comment generation, uniquely featuring complex analytical queries with expert-refined annotations.
\item We propose SQL-Commenter, a new method combining CPT, SFT, and DPO for SQL code comment generation, achieving state-of-the-art performance through automatic metrics and human evaluation.
\end{itemize}

We have made all source code and datasets publicly available at \url{https://zenodo.org/records/18626983}



\section{Background and Motivation}
\label{sec:background}

SQL is fundamental to data management, but growing business complexity leads to intricate queries involving joins, subqueries, and CTEs. Understanding these queries is a significant challenge that hinders maintenance, debugging, and collaboration. While high-quality comments are crucial for clarity, they are often missing or outdated, motivating the need for automated solutions to generate them.

\subsection{Problem Statement}
\label{subsec:problem_statement}

We propose an automated approach for SQL query comment generation. Given an SQL query \(x\), our system generates a descriptive comment \(\hat{y}\) that accurately captures the query's functionality and business logic. The task is formalized as finding an optimal mapping function \(f: X \rightarrow Y\) where \(X\) represents the space of SQL queries and \(Y\) represents natural language comments. Our goal is to train a model \(f\) that produces comments that are not only factually correct but also semantically precise and contextually relevant.

\subsection{Motivations}
\label{subsec:motivations}


In this section, we analyze the motivations behind our research on SQL query comment generation.



\begin{figure}[htbp]
\centerline{\includegraphics[width=0.45\textwidth,height=0.4\textheight]{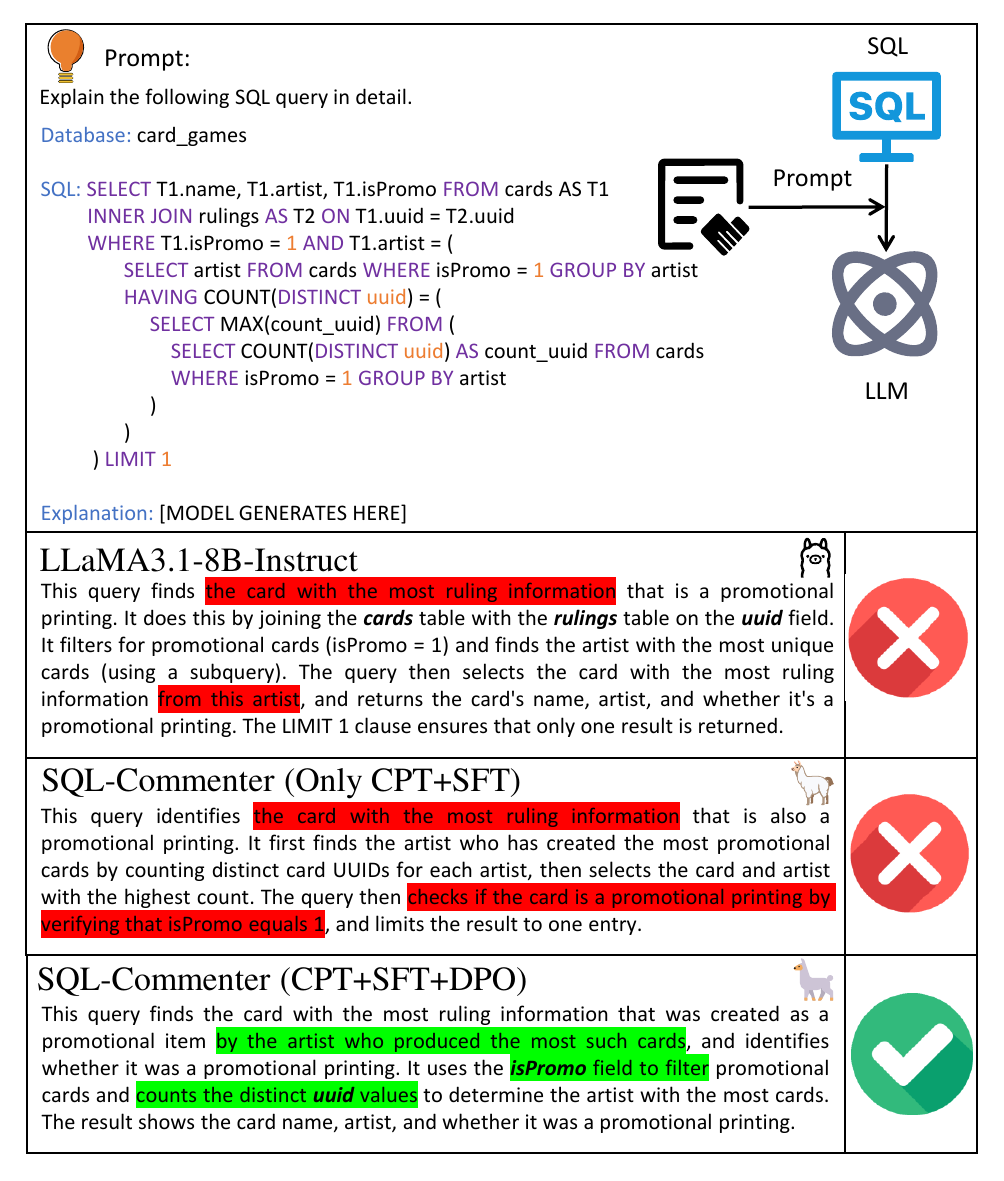}}
\caption{The Motivation of our SQL-Commenter.}
\label{tab: motivation}
\end{figure}

\textbf{Motivation 1: Limitations of Existing Datasets and Naive Conversion Methods for SQL-to-Text.}
A primary challenge in developing robust SQL comment generators is the scarcity of high-quality, large-scale SQL-to-Text datasets. Most existing resources, such as the widely-used Spider~\cite{yu2018spider} and BIRD~\cite{li2023can} benchmarks, were designed for the Text-to-SQL task. Consequently, their natural language annotations are typically short, high-level questions that describe user \textit{intent} (e.g., "Show me the names of all singers sorted by age.") rather than providing a technical \textit{explanation} of the SQL query's execution logic.

Simply inverting these datasets for SQL-to-Text, a common practice in prior work like HeSQL-Net~\cite{zhang2025hesqlnet}, is a naive approach with significant drawbacks. Using the original user question as the ground truth for a generated comment results in annotations that are unnatural and lack technical depth. They fail to explain \textit{how} the query works---omitting details about join types, filtering logic, aggregation methods, and subquery purposes. This creates a fundamental mismatch between the training objective and the goal of generating truly informative, explanatory comments.

To overcome this critical limitation, we propose a novel data transformation methodology. Instead of merely reusing the original questions, we leverage the original Text-to-SQL pairs from benchmarks like Spider and Bird to generate new, high-quality ground truths. Our automated pipeline constructs a detailed prompt for each sample, including the SQL query, the original natural language question, and relevant database schema information (for the BIRD dataset, this also includes the "evidence" text). By employing a few-shot prompting strategy with carefully crafted examples, we guide the DeepSeek-V3.1 model (invoked via its efficient non-thinking mode) to generate a detailed, technically sound comment that explains the query’s step-by-step functionality. This process transforms the original intent-based questions into rich, explanatory ground truths. All generated comments are subsequently verified and refined by human experts to ensure accuracy and quality, resulting in a dataset that is truly tailored for training and evaluating high-fidelity SQL comment generation models.

\renewcommand{\theadfont}{\normalsize\bfseries}

\begin{table}[h!]
\centering
\caption{Comparison of Comment Quality. "Ours (SFT)" refers to SQL-Commenter after CPT+SFT, and "Ours (DPO)" includes the final DPO step.}
\label{tab:motivation_summary}
\begin{tabularx}{\columnwidth}{>{\raggedright\arraybackslash}X c c c}
\toprule
\thead{Semantic Feature} & \thead{LLaMA3.1-8B \\ Instruct} & \thead{Ours \\ (SFT)} & \thead{Ours \\ (DPO)} \\
\midrule
Identifies basic JOIN operation & \cmark & \cmark & \cmark \\
\addlinespace
Identifies `isPromo = 1` filter & \cmark & \cmark & \cmark \\
\addlinespace
Identifies subquery's purpose (top promo artist) & \xmark & \cmark & \cmark \\
\addlinespace
Identifies \texttt{rulings} JOIN function (linking, not aggregation) & \xmark & \xmark & \cmark \\
\addlinespace
Identifies overall query intent (find card from top artist) & \xmark & \xmark & \cmark \\

\bottomrule
\end{tabularx}
\end{table}

\textbf{Motivation 2: Limitations in Comment Quality of LLM-based SQL Comment Generation Methods.}

As demonstrated in Figure \ref{tab: motivation} of the paper and summarized in Table \ref{tab:motivation_summary}, even with domain-specific training data, Large Language Models (LLMs) often struggle to grasp the deep, compositional semantics of complex SQL queries.

General-purpose LLMs (e.g., LLaMA-3.1-8B-Instruct) tend to produce comments that are either vague or factually incorrect. As shown in Table \ref{tab:motivation_summary}, this model fails to decipher the logic of the complex, nested subquery, which is designed to identify the artist who has produced the most promotional cards. It incorrectly infers that the query's goal is to find the card with the "most ruling information," resulting in a superficial and misleading summary.

Although a specialized LLM, such as SQL-Commenter trained only with CPT and SFT, shows significant improvement, it still exhibits subtle but critical flaws. It correctly identifies the subquery's purpose but fails to understand the interaction between query components. Its explanation contains a crucial semantic error: it claims the final query selects the card with the "most ruling information *from this artist*," which is incorrect as no aggregation is applied to the `rulings` table. It describes the query's components accurately but fails to reason about their interaction to understand the query's true, holistic intent. This limitation underscores the need for a more advanced training methodology.

To bridge this semantic gap, we introduce Direct Preference Optimization (DPO). As seen in the final column of Table \ref{tab:motivation_summary}, the full SQL-Commenter model correctly interprets all aspects of the query. DPO achieves this through two key aspects:
\begin{enumerate}
\item \textbf{Fine-grained Semantics Learning:} DPO trains on preference pairs where explanations reflecting holistic intent are favored over those merely describing components. This process teaches the model the deep semantic rules of complex query constructs.
\item \textbf{Context-dependent Quality Assessment:} The preference mechanism also captures contextual nuance. By training on pairs that reflect different user needs (e.g., technical vs. business logic summaries), the model learns to generate comments tailored to the target audience.
\end{enumerate}

DPO thus addresses the deep semantic limitations of CPT and SFT-only models, enabling the generation of higher-quality, context-aware SQL comments.

\section{Approach}

SQL-Commenter consists of three key stages: Continual Pre-Training, Supervised Fine-Tuning, and Direct Preference Optimization. Open-source SQL Query Collection, SFT and DPO data construction, and Evaluation of SQL Comments further support these core stages. We selected LLaMA-3.1-8B as the base model due to its open-source characteristics, optimal parameter size, and excellent capability for fine-tuning \cite{alrashedy2023language, lu2023llama, yu2024smart, tang2025breaking}.

\subsection{Open-source SQL Query Collection}

\subsubsection{Continual Pre-training}

For our Continual Pre-training dataset, we integrated SQL-related data from multiple sources: (1) The official documentation for major database systems (e.g., PostgreSQL, MySQL), containing detailed function descriptions and query examples; (2) Stack Overflow question-answer pairs tagged with "sql", "mysql", "t-sql", and other database-related keywords; (3) Highly-starred data analysis and backend development projects from GitHub that contain a large number of SQL queries; (4) Curated SQL query discussions from platforms like DBA Stack Exchange. We extracted a total of approximately 1,200,000 SQL queries from these sources. To ensure uniqueness, queries were filtered using a Jaccard Index similarity threshold of 0.9, eliminating those with over 90\% token similarity. After deduplication, around 850,000 unique SQL queries remained. The threshold of 0.9 was adopted following prior work, which used this value to identify and remove near-duplicate code snippets.

\subsubsection{Supervised Fine-Tuning and Direct Preference Optimization}

\textbf{For SFT dataset}, we combined data from two primary sources: the well-known text-to-SQL benchmark, Spider, and more complex queries sourced from the Bird benchmark and public code repositories. The Spider dataset \cite{yu2018spider} provided 7,215 annotated SQL queries with high-quality comments. To augment this, we incorporated an additional 7,856 annotated complex SQL queries from Bird dataset \cite{li2023can}, focusing on diverse use cases such as multi-table joins, window functions, and complex aggregations. After quality control and deduplication, our final SFT dataset contained approximately 15,071 high-quality $\langle$SQL Query, Comment$\rangle$ pairs. \textbf{For DPO dataset}, we constructed paired data comprising preferred and rejected comments for the same SQL queries. We constructed 1,849 preference pairs from the Spider dataset. Additionally, we created 1,167 preference pairs from Bird dataset. We ensured there was no overlap between the DPO dataset and the SFT dataset. In total, our DPO dataset consisted of 3,016 preference pairs covering various SQL usage scenarios from simple lookups to complex analytical queries.

\subsection{SFT and DPO Data Construction}

\begin{figure*}[htbp]
\centerline{\includegraphics[width=0.85\textwidth,height=0.42\textheight]{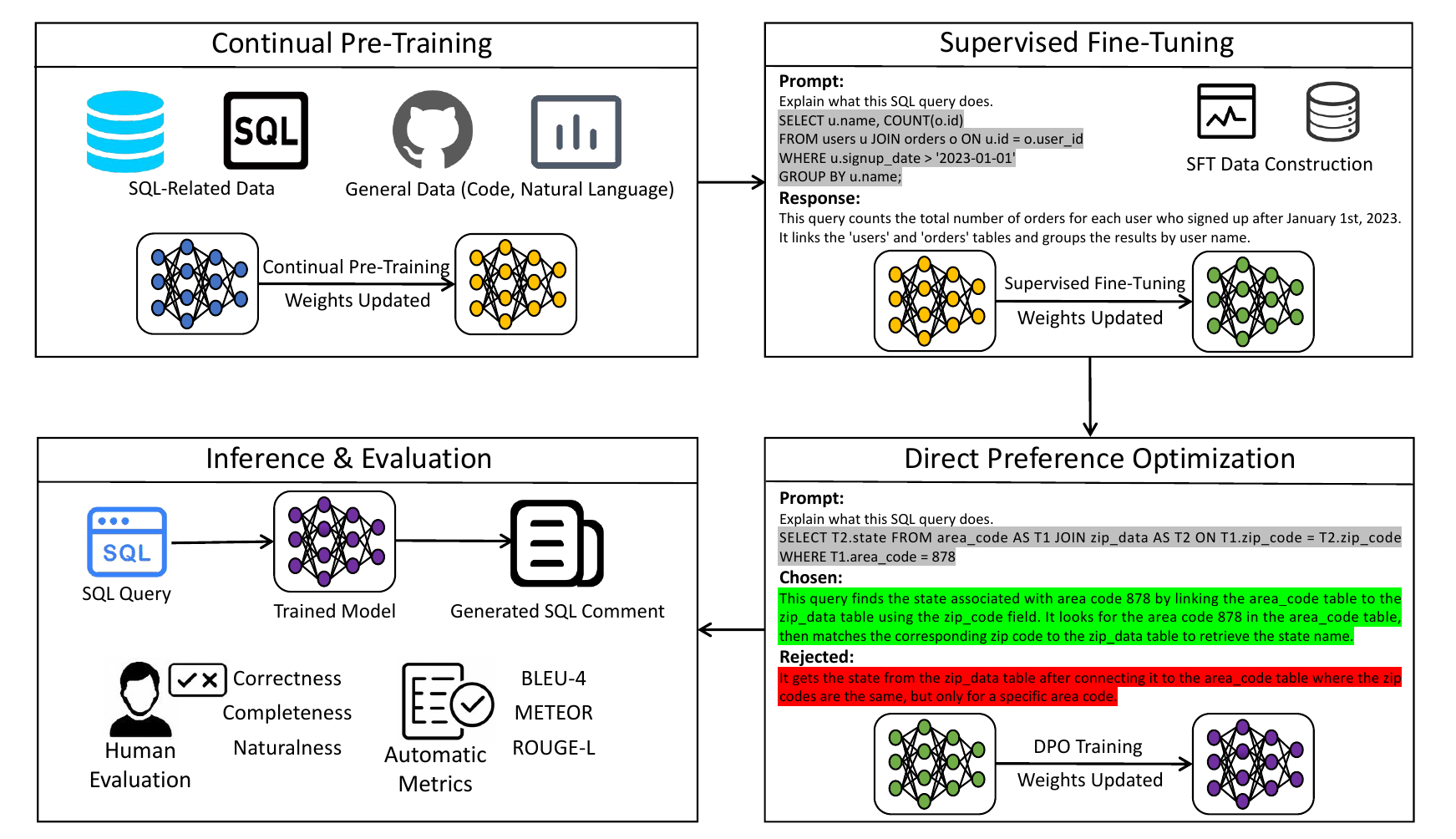}}
\caption{The Overview of our SQL-Commenter.}
\label{tab: overview}
\end{figure*}

\subsubsection{SFT Data Construction}
Our SFT data construction follows a two-stage, model-plus-human pipeline.

First, we perform automated initial annotation generation. We leverage the original Text-to-SQL data pairs from the Spider and BIRD benchmarks. For each pair, we create a structured prompt that includes the SQL query, the original natural language question, and database context (for BIRD, we also include the provided "evidence" text). Using a few-shot strategy with curated examples, we prompt the DeepSeek-V3.1 model via its non-thinking mode to generate a concise, technically precise explanation of the SQL query. This approach ensures the initial comments are not just high-level summaries but detailed technical descriptions.

Second, these machine-generated comments undergo expert review and refinement. A team of 20 data analysts and database administrators, each with over 5 years of SQL experience, meticulously verifies the accuracy and completeness of every annotation. The experts correct any logical inaccuracies, improve explanations of complex components like joins, window functions, and aggregations, and ensure the final language is clear and natural. This rigorous human-in-the-loop process transforms the initial LLM outputs into a high-fidelity dataset of 15,071 ⟨SQL Query, Comment⟩ pairs, forming a robust foundation for supervised fine-tuning.

\subsubsection{DPO Data Construction}

To construct our DPO dataset, we generated 3,016 new preference pairs, ensuring no data overlap with our SFT training set to maintain methodological integrity. The "chosen" (preferred) responses within these pairs were created using the same high-quality annotation methodology developed for our SFT data, thereby establishing a consistent standard of excellence. The corresponding "rejected" comments were produced by an automated pipeline that leverages a powerful LLM (DeepSeek-V3.1) guided by a dynamically constructed, multi-strategy prompt. This prompt first sets the task and provides the SQL query and the "chosen" comment as context. The core of this framework is a multi-strategy negative sampling mechanism where, for each sample, the system randomly selects one of eight predefined negative strategies: superficial, incomplete, technical\_errors, overly\_verbose, vague\_and\_unclear, wrong\_emphasis, poor\_structure, and misunderstand\_purpose. A detailed instruction corresponding to the selected strategy is then dynamically injected into the prompt—for instance, explicitly telling the LLM to ``Generate an INCOMPLETE explanation that feels rushed...'' for the incomplete strategy. Finally, a strict directive (``Generate ONLY the poor explanation...'') ensures clean data extraction. This structured approach systematically creates a clear but reasonable quality gap, yielding a diverse and challenging set of preference pairs for effective DPO training.

\subsection{Continual Pre-Training}

To develop SQL-Commenter, we continually pre-train LLaMA-3.1-8B for two epochs on SQL-related data and one epoch on natural language related data. This mixed training in natural language and code offers benefits for a wide range of tasks in both domains. We optimize the language modeling objective that is widely used in prior pre-trained language models like GPT and LLaMA. Specifically, given a sequence $x$ consisting of $n$ tokens, denoted as $t_0, t_1, t_2, ..., t_{n-1}$, our objective is to maximize the likelihood of the entire sequence. This is achieved by calculating the product of the conditional probabilities for each token:

\begin{equation}
p(x) = \prod_{i=1}^{n-1} p(t_i|t_1, t_2, ...t_{i-1})
\end{equation}

Pre-training Dataset [Total 2.18B tokens]. To build a comprehensive and efficient pre-training dataset, we curated high-quality data from two distinct, well-established sources:

SQL-related data [1.91B tokens]. To enhance the model's fundamental understanding of SQL syntax, semantics, and common patterns, we utilized the SQL portion of The Stack v1.2 \cite{kocetkov2022the}. While our data selection is based on the StarCoder project \cite{li2023starcoder}, we implemented an additional rigorous deduplication step to improve data quality and diversity. Specifically, we calculated the Jaccard similarity between all text and discarded those with a similarity score of 0.8 or higher, retaining only text with lower similarity. This refined corpus, comprising 634,807 instances and totaling 1.91B tokens, provides a robust foundation in the SQL domain, covering everything from simple queries to complex analytical scripts.

General domain data [272M tokens]. To maintain and bolster the model's general reasoning, instruction-following, and multilingual abilities, we incorporated a diverse, high-quality corpus of general-purpose data. This component, sourced from the pre-training mixture developed for the MaP-Neo model series \cite{zhang2024map}, consists of 300,000 instances, totaling 272M tokens. It includes a balanced collection of natural language (English and Chinese), general-purpose code (Python, Java, C++), and mathematical data, which is crucial for preventing catastrophic forgetting and enhancing the model's overall intelligence.

\subsection{Supervised Fine-Tuning}

In the Supervised Fine-Tuning stage, we refine the large language model to produce high-quality comments for SQL queries. For this purpose, we employ a token-level negative log-likelihood loss optimized over the dataset:

\begin{equation}
\mathcal{L}{\mathrm{SFT}} = -\frac{1}{|\mathcal{D}{\mathrm{exp}}|} \sum_{(x, y) \in \mathcal{D}{\mathrm{exp}}} \sum{t=1}^{|y|} \log P_{\theta}(y_t \mid x, y_{<t})
\end{equation}

Here, $x$ is the input SQL query, $y$ the target comment sequence, and $\theta$ denotes all trainable parameters of the LLM.

This objective enables the model to learn to generate comments that accurately and fluently explain the functional and contextual aspects of SQL queries. The prior domain-specific knowledge acquired during continual pre-training helps the model better understand SQL syntax and semantics, thus supporting the generation of detailed and correct comments.

\subsection{Direct Preference Optimization}

In the final stage, we employ Direct Preference Optimization (DPO) to align the LLM's outputs with the preferences of human developers for high-quality SQL query comments. DPO simplifies preference learning by directly optimizing the model without requiring explicit reward modeling or reinforcement learning.

The core of DPO is based on the insight that the optimal policy $\pi^*$ for a reward function $r^*$ under a KL-constrained optimization objective can be expressed as:
\begin{equation}
\pi^*(y|x) = \frac{1}{Z(x)} \pi_{\text{ref}}(y|x) \exp\left(\frac{1}{\beta} r^*(x,y)\right)
\end{equation}
where $Z(x)$ is a normalization factor, $\pi_{\text{ref}}$ is a reference policy, and $\beta$ is a temperature parameter. In our SQL comment generation scenario:
\begin{itemize}
\item $x$: represents the input SQL query
\item $y$: represents the LLM-generated comment
\item $\pi_{\text{ref}}$: the reference policy, typically the initially fine-tuned model trained in the SFT stage
\item $\pi^*$: the final optimized model that generates comments aligning with developer preferences
\end{itemize}

By rearranging this equation, we can express the reward function in terms of the optimal policy:
\begin{equation}
r^*(x,y) = \beta \log\frac{\pi^*(y|x)}{\pi_{\text{ref}}(y|x)} + \beta \log Z(x)
\end{equation}

This allows us to reformulate the Bradley-Terry preference model in terms of policies rather than rewards:
\begin{equation}
p^*(y_p \succ y_n \mid x) = \sigma\left(\beta \log\frac{\pi^*(y_p|x)}{\pi_{\text{ref}}(y_p|x)} - \beta \log\frac{\pi^*(y_n|x)}{\pi_{\text{ref}}(y_n|x)}\right)
\end{equation}
where $\sigma$ is the logistic function, and $y_p$ and $y_n$ represent the preferred and non-preferred comment outputs.

The DPO loss function is then expressed as:
\begin{equation} \label{eq:dpo_loss}
\begin{split}
\mathcal{L}_{\text{DPO}}(\pi_\theta; \pi_{\text{ref}}) = -\mathbb{E}_{(x, y_p, y_n) \sim \mathcal{D}} \bigg[ \log \sigma\bigg( & \beta \log\frac{\pi_\theta(y_p|x)}{\pi_{\text{ref}}(y_p|x)} \\
& - \beta \log\frac{\pi_\theta(y_n|x)}{\pi_{\text{ref}}(y_n|x)}\bigg)\bigg]
\end{split}
\end{equation}
where $(x, y_p, y_n)$ are triples from our curated dataset $\mathcal{D}$, representing an input SQL query, a preferred output, and a non-preferred output. By minimizing this loss, we directly optimize the policy model $\pi_\theta$ to align with the preferences of data analysts and database developers, enabling SQL-Commenter to generate comments that are accurate, comprehensive, and developer-friendly.

\subsection{Human Evaluation of SQL Comments}


For human evaluation, we recruited six SQL experts, each with over 5 years of experience and strong English proficiency. We selected 300 SQL queries (200 from Spider, 100 from Bird) using stratified sampling to reflect the original distribution of query difficulty. To establish inter-evaluator agreement, all six experts first rated a shared calibration set. This set comprised 30 of the selected queries (20 from Spider, 10 from Bird), and for each query, the experts evaluated the comments generated by our model and the two baselines. We calculated Fleiss’ Kappa on these ratings, achieving a score of 0.71, which indicates substantial agreement. Following calibration, the remaining 270 queries were partitioned into three subsets of 90. Each of our three expert pairs was assigned one subset, for which they rated the comments from all three models. This design ensured that every primary evaluation comment was independently assessed by two experts. All comments were rated on a 1–4 Likert scale for three metrics: Correctness (whether the comment accurately reflects the query's functionality and logic), Completeness (whether it covers all important elements, such as joins, filters, and aggregations), and Naturalness (whether its language is fluent and easy to understand). The final score for each metric on a given comment was calculated as the average of the two experts' ratings. In cases where the average resulted in a non-integer (e.g., 3.5), the score was rounded down to the next lowest integer.

\section{Experiments}

\subsection{Research Questions}
To evaluate our proposed SQL-Commenter approach, we conduct experiments to answer the following research questions:
\begin{itemize}
    \item[\textbf{RQ1:}] How does SQL-Commenter perform in the SQL comment generation task compared to existing state-of-the-art methods?
    \item[\textbf{RQ2:}] How effective are the individual components of our SQL-Commenter?
    \item[\textbf{RQ3:}] How effective are the comments generated by SQL-Commenter in terms of Correctness, Completeness, and Naturalness, as measured by human evaluation?
    \item[\textbf{RQ4:}] What are the primary failure modes of SQL-Commenter, and what do they reveal about its potential avenues for future improvement?

\end{itemize}

\subsection{Dataset}

Our data pipeline involves several stages, starting with \textbf{CPT}. For this, we built a 2.18B token dataset composed of two parts to ensure both domain expertise and general capability: a 1.91B token domain-specific portion derived from the SQL subset of The Stack v1.2 \cite{kocetkov2022the} as curated for the StarCoder project \cite{li2023starcoder}, and a 272M token supplement from the MaP-Neo project's \cite{zhang2024map} high-quality bilingual corpus for general reasoning. Next, for \textbf{SFT}, we combined data from the Spider \cite{yu2018spider} and Bird \cite{li2023can} benchmarks, which, after quality control and expert refinement, resulted in a final dataset of 15,071 high-quality ⟨SQL Query, Comment⟩ pairs totaling 2.94M tokens. For \textbf{DPO}, we created a dataset of 3,016 preference pairs (1.24M tokens), where each pair comprises a preferred and a rejected comment for the same SQL query; the queries were sourced from the Spider and Bird training splits and were kept entirely distinct from the SFT set to prevent data overlap. Finally, for \textbf{Evaluation}, we repurposed original Text-to-SQL pairs to create three SQL-to-Text test sets: spider\_dev (1,034 samples), spider\_test (2,147 samples), and bird\_dev (1,534 samples). The bird\_test set was inaccessible because the original BIRD benchmark keeps it private to prevent data leakage. For human evaluation, due to resource constraints, we randomly sampled 200 queries from spider\_test and 100 from bird\_dev.


\subsection{Baselines}
We compare SQL-Commenter against a wide range of recent and powerful LLMs, which can be categorized as follows:

\begin{itemize}
    \item \textbf{General-purpose LLMs:} Models from the LLaMA series \cite{grattafiori2024llama} (Llama-3.1-8B-Instruct, Llama-3.2-1B/3B-Instruct) and the Qwen series \cite{yang2025qwen3} (Qwen2.5-3B/7B/14B/32B-Instruct and Qwen3-8B/14B).
    \item \textbf{Code-specialized LLMs:} Models from the DeepSeek-Coder series\cite{guo2024deepseek} (Deepseek-Coder-6.7B/33B-Instruct), the Qwen2.5-Coder series \cite{hui2024qwen2} (Qwen2.5-Coder-1.5B/3B/7B/14B-Instruct), and the CodeLLaMA series \cite{roziere2023code} (CodeLlama-7b/13b-Instruct).
    \item \textbf{Distilled LLMs:} Models from the DeepSeek-R1-Distill series \cite{guo2025deepseek}, including DeepSeek-R1-Distill-Qwen-1.5B/7B/14B/32B and DeepSeek-R1-Distill-Llama-8B.
\end{itemize}

We exclude earlier graph-based models (Graph2Seq \cite{ma2021relation}, HeSQLNet \cite{zhang2025hesqlnet}) as they are designed for the "SQL-to-Question" task, which generates concise, high-level questions. Their graph-to-sequence architectures are not well-suited to generalize to the generation of longer, detailed technical explanations required by our work. Consequently, their outputs lack the necessary technical depth, rendering a direct comparison unsuitable for this study.

\subsection{Metrics}
We evaluated our model from two dimensions: automatic metrics and human evaluation. For automatic evaluation, we used \textbf{BLEU-4} (n-gram precision), \textbf{METEOR} (precision/recall with synonyms and stemming), and \textbf{ROUGE-L} (longest common subsequence). For human evaluation, we employed three key metrics: \textbf{Correctness}, \textbf{Completeness}, and \textbf{Naturalness}.

\subsection{Implementation Details}
All models (CPT, SFT, and DPO) were trained using LlamaFactory \cite{zheng2024llamafactory} and DeepSpeed with fp16 precision. We used the AdamW optimizer ($\beta = (0.9, 0.99), \epsilon = 10^{-8}$) with full parameter tuning on 8 NVIDIA H800 GPUs (80GB memory each). For CPT, we used a batch size of 64, 16 gradient accumulation steps, and trained for 3 epochs with a learning rate of $1 \times 10^{-5}$. For SFT, the batch size was 8, with 8 gradient accumulation steps for 3 epochs. For DPO, the batch size was 8, with 4 gradient accumulation steps for 3 epoch. All learning rates used a cosine decay schedule with no warmup. Evaluation was performed using greedy decoding (temperature = 0). For all baseline LLMs, we used a few-shot (3-shot) prompting strategy, with examples drawn from the respective training sets of Spider and Bird.


\begin{table*}[htbp]
\centering
\caption{Performance comparison of SQL-Commenter with baseline models on the Spider and Bird datasets. We report BLEU-4 (B-4), METEOR (M), and ROUGE-L (R-L) scores. All values are in percentages (\%). Best results are in \textbf{bold}.}
\label{tab:full_comparison}
\renewcommand{\arraystretch}{1.1} 
\setlength{\tabcolsep}{4pt} 

\begin{tabular}{ll ccc ccc ccc}
\toprule
\multirow{2}{*}{\textbf{Method Type}} & \multirow{2}{*}{\textbf{Method Name}} & \multicolumn{3}{c}{\textbf{Spider (dev)}} & \multicolumn{3}{c}{\textbf{Spider (test)}} & \multicolumn{3}{c}{\textbf{Bird (dev)}} \\
\cmidrule(lr){3-5} \cmidrule(lr){6-8} \cmidrule(lr){9-11}
& & B-4 & M & R-L & B-4 & M & R-L & B-4 & M & R-L \\
\midrule
\multirow{5}{*}{DeepSeek} 
& DeepSeek-R1-Distill-Qwen-1.5B   & 21.30 & 46.55 & 38.06 & 19.96 & 45.28 & 37.20 & 18.70 & 43.03 & 36.28 \\
& DeepSeek-R1-Distill-Qwen-7B   & 23.71 & 46.16 & 42.13 & 22.67 & 45.40 & 41.25 & 21.93 & 44.75 & 41.18 \\
& DeepSeek-R1-Distill-Llama-8B  & 26.14 & 50.55 & 42.66 & 25.49 & 49.91 & 42.46 & 21.98 & 46.39 & 40.10 \\
& DeepSeek-R1-Distill-Qwen-14B  & 23.87 & 48.88 & 41.52 & 23.87 & 48.54 & 41.31 & 23.38 & 48.79 & 40.64 \\
& DeepSeek-R1-Distill-Qwen-32B  & 26.25 & 51.40 & 42.93 & 25.81 & 50.87 & 42.50 & 24.96 & 49.92 & 41.77 \\
\midrule
\multirow{2}{*}{DeepSeek-Coder} 
& Deepseek-Coder-6.7B-Instruct  & 24.10 & 48.72 & 40.99 & 23.25 & 47.30 & 40.13 & 21.77 & 46.62 & 39.00 \\
& Deepseek-Coder-33B-Instruct & 23.89 & 49.43 & 41.72 & 23.71 & 48.48 & 41.50 & 24.43 & 48.75 & 41.97 \\
\midrule
\multirow{4}{*}{Qwen2.5} 
& Qwen2.5-3B-Instruct           & 6.55 & 18.65 & 25.84 & 6.46 & 18.60 & 25.53 & 8.83 & 22.85 & 28.88 \\
& Qwen2.5-7B-Instruct           & 10.90 & 33.52 & 29.52 & 10.32 & 32.49 & 29.19 & 11.22 & 33.10 & 31.42 \\
& Qwen2.5-14B-Instruct          & 11.68 & 30.91 & 31.47 & 11.45 & 30.38 & 31.10 & 11.76 & 33.10 & 31.43 \\
& Qwen2.5-32B-Instruct          & 10.68 & 27.89 & 31.78 & 10.38 & 27.29 & 31.82 & 13.38 & 39.68 & 32.68 \\
\midrule
\multirow{4}{*}{Qwen2.5-Coder} 
& Qwen2.5-Coder-1.5B-Instruct   & 7.30 & 29.34 & 22.08 & 7.47 & 29.50 & 22.23 & 10.17 & 38.90 & 23.21 \\
& Qwen2.5-Coder-3B-Instruct     & 15.97 & 47.29 & 31.37 & 15.63 & 46.73 & 30.74 & 13.00 & 45.55 & 29.16 \\
& Qwen2.5-Coder-7B-Instruct     & 22.61 & 50.93 & 39.68 & 21.97 & 49.94 & 38.75 & 22.37 & 52.20 & 39.25 \\
& Qwen2.5-Coder-14B-Instruct    & 24.88 & 54.89 & 42.22 & 24.31 & 53.81 & 41.38 & 22.17 & 50.51 & 39.73 \\
\midrule
\multirow{2}{*}{Qwen3} 
& Qwen3-8B                      & 24.53 & 51.65 & 42.24 & 25.39 & 51.44 & 42.91 & 21.33 & 48.93 & 38.23 \\
& Qwen3-14B                     & 27.93 & 53.57 & 43.76 & 26.89 & 51.96 & 44.45 & 25.73 & 51.54 & 42.50 \\
\midrule
\multirow{3}{*}{LLaMA} 
& Llama-3.2-1B-Instruct         & 14.03 & 39.55 & 31.35 & 13.44 & 38.30 & 30.60 & 20.15 & 45.52 & 37.56 \\
& Llama-3.2-3B-Instruct         & 16.80 & 39.77 & 33.84 & 16.15 & 38.84 & 33.12 & 19.31 & 43.03 & 37.64 \\
& Llama-3.1-8B-Instruct         & 18.43 & 42.45 & 34.74 & 17.24 & 40.89 & 33.58 & 22.26 & 47.26 & 39.79 \\
\midrule
\multirow{2}{*}{CodeLLaMA} 
& CodeLlama-7b-Instruct         & 24.53 & 49.61 & 41.82 & 23.77 & 48.85 & 41.08 & 21.13 & 46.37 & 39.22 \\
& CodeLlama-13b-Instruct        & 26.03 & 50.01 & 43.61 & 25.44 & 49.29 & 43.11 & 22.64 & 47.18 & 40.24 \\
\midrule
\textbf{Our Method} & \textbf{SQL-Commenter} & \textbf{36.95} & \textbf{58.37} & \textbf{57.17} & \textbf{36.37} & \textbf{57.76} & \textbf{56.48} & \textbf{35.09} & \textbf{55.91} & \textbf{56.74} \\
\bottomrule
\end{tabular}
\end{table*}

\begin{table*}[htbp]
\centering
\caption{Ablation study of SQL-Commenter's components on the Spider and Bird datasets. We report BLEU-4 (B-4), METEOR (M), and ROUGE-L (R-L) scores (\%). Best results are in \textbf{bold}.}
\label{tab:ablation_full}
\renewcommand{\arraystretch}{1.2}
\setlength{\tabcolsep}{5pt}

\begin{tabular}{l ccc ccc ccc}
\toprule
\multirow{2}{*}{\textbf{Configuration}} & \multicolumn{3}{c}{\textbf{Spider (dev)}} & \multicolumn{3}{c}{\textbf{Spider (test)}} & \multicolumn{3}{c}{\textbf{Bird (dev)}} \\
\cmidrule(lr){2-4} \cmidrule(lr){5-7} \cmidrule(lr){8-10}
& B-4 & M & R-L & B-4 & M & R-L & B-4 & M & R-L \\
\midrule
\textbf{SQL-Commenter (Full Model)} & \textbf{36.95} & \textbf{58.37} & \textbf{57.17} & \textbf{36.37} & \textbf{57.76} & \textbf{56.48} & \textbf{35.09} & \textbf{55.91} & \textbf{56.74} \\
\midrule
w/o DPO (CPT + SFT)                 & 27.55 & 55.72 & 42.63 & 27.19 & 55.24 & 42.64 & 23.76 & 51.36 & 39.69 \\
w/o CPT (SFT + DPO)                 & 14.85 & 40.23 & 30.16 & 14.08 & 39.67 & 29.47 & 14.32 & 40.26 & 30.28 \\
w/o SFT (CPT + DPO)                 & 0.28 & 11.08 & 13.43 & 0.18 & 8.66 & 11.51 & 0.63 & 12.74 & 16.89 \\
\bottomrule
\end{tabular}
\end{table*}

\begin{table*}[t!]
\centering
\caption{Human evaluation ratings of Correctness, Completeness, and Naturalness. Each cell shows the count of ratings for scores 1 (worst) to 4 (best). Total samples: 200 for Spider, 100 for Bird. DeepSeek-32B refers to DeepSeek-R1-Distill-Qwen-32B.}
\label{tab:human_eval}
\begin{tabular}{ll cccc cccc cccc}
\toprule
\multirow{2}{*}{\textbf{Dataset}} & \multirow{2}{*}{\textbf{Model}} & \multicolumn{4}{c}{\textbf{Correctness}} & \multicolumn{4}{c}{\textbf{Completeness}} & \multicolumn{4}{c}{\textbf{Naturalness}} \\
\cmidrule(lr){3-6} \cmidrule(lr){7-10} \cmidrule(lr){11-14}
& & \textbf{1} & \textbf{2} & \textbf{3} & \textbf{4} & \textbf{1} & \textbf{2} & \textbf{3} & \textbf{4} & \textbf{1} & \textbf{2} & \textbf{3} & \textbf{4} \\
\midrule
\multirow{3}{*}{Spider} & DeepSeek-32B & 24 & 41 & 78 & 57 & 29 & 48 & 81 & 42 & 11 & 29 & 92 & 68 \\
& Qwen3-14B & 19 & 42 & 81 & 58 & 26 & 46 & 84 & 44 & 9 & 27 & 96 & 68 \\
& \textbf{Ours} & \textbf{8} & \textbf{17} & \textbf{83} & \textbf{92} & \textbf{11} & \textbf{23} & \textbf{86} & \textbf{80} & \textbf{4} & \textbf{11} & \textbf{83} & \textbf{102} \\
\midrule
\multirow{3}{*}{Bird} & DeepSeek-32B & 19 & 28 & 33 & 20 & 21 & 33 & 34 & 12 & 8 & 17 & 49 & 26 \\
& Qwen3-14B & 16 & 26 & 38 & 20 & 18 & 31 & 39 & 12 & 7 & 14 & 51 & 28 \\
& \textbf{Ours} & \textbf{8} & \textbf{16} & \textbf{44} & \textbf{32} & \textbf{11} & \textbf{19} & \textbf{46} & \textbf{24} & \textbf{4} & \textbf{9} & \textbf{43} & \textbf{44} \\
\bottomrule
\end{tabular}
\end{table*}


\subsection{Experimental Results}

\subsubsection{RQ1: Overall Performance}



To answer this question, we compared SQL-Commenter's performance against a comprehensive set of baseline models on the Spider and Bird benchmarks. As shown in Table~\ref{tab:full_comparison}, SQL-Commenter achieves favorable results compared to other methods across all datasets and metrics.

SQL-Commenter shows strong performance on the Spider benchmark. On the widely-used Spider test set, it achieves a BLEU-4 score of 36.37\%, substantially outperforming the 26.89\% from the strongest baseline, Qwen3-14B. This performance advantage extends to other metrics, with our model scoring 57.76\% on METEOR and 56.48\% on ROUGE-L. A similar trend is observed on the Spider dev set, where our model achieves a BLEU-4 of 36.95\%, again clearly ahead of the baseline's performance. The consistent gap across both sets suggests an enhanced ability of our model to capture semantic meaning and generate fluent comments.

On the more challenging Bird dev set, which features complex, real-world analytical queries, SQL-Commenter continues to perform strongly. It achieves a BLEU-4 score of 35.09\%, maintaining a significant lead over Qwen3-14B's score of 25.73\%. This result, alongside high scores in METEOR (55.91\%) and ROUGE-L (56.74\%), indicates the robustness of our three-stage training approach in handling SQL logic where other LLMs may struggle.

\begin{tcolorbox}
\textbf{[RQ1]: SQL-Commenter outperformed existing state-of-the-art methods across all main automatic evaluation metrics (BLEU-4, METEOR, and ROUGE-L) in SQL comment generation.}
\end{tcolorbox}

\subsubsection{RQ2: Ablation Study}
To validate the contribution of each component in our methodology, we conducted a thorough ablation study. 
The results, presented in Table~\ref{tab:ablation_full}, confirm that while all components are indispensable for overall functionality, Direct Preference Optimization (DPO) plays the most critical role in elevating the model's output to expert-level quality.

The significance of \textbf{DPO} (\texttt{w/o DPO}), our core contribution for quality refinement, is immediately apparent. 
Removing it causes a severe and consistent performance drop across all metrics and datasets. 
For instance, BLEU-4 scores decrease by \textbf{9.40} on \textbf{Spider dev}, \textbf{9.18} on \textbf{Spider test}, and a substantial \textbf{11.33} on \textbf{Bird dev}. 
The core insight here lies not just in the score degradation, but in the fundamental learning paradigm. 
Supervised Fine-Tuning (SFT) trains the model to mimic a single ground-truth target, effectively learning \textit{what} a correct comment looks like. 
DPO, in contrast, learns from a preference signal over pairs of responses (chosen vs. rejected). This paradigm compels the model to discern the underlying principles of what constitutes a high-quality response, rather than merely replicating a single valid example.
It learns to optimize for attributes like clarity, conciseness, and logical coherence that differentiate expert-level comments. 
This marks the transition from a model capable of generating functionally correct text to one that produces refined, expert-quality commentary.

In contrast, the other components serve as essential foundations. 
The performance collapse observed when removing \textbf{Continued Pre-training} (\texttt{w/o CPT}) underscores its role as the foundational stage for domain-specific knowledge acquisition. 
CPT endows the model with a comprehensive understanding of SQL's syntactic and semantic structures. 
Without this pre-training, the model lacks the prerequisite domain knowledge to interpret the input queries meaningfully, rendering the subsequent fine-tuning stages on the specific task largely ineffective. 
This is evidenced by the dramatic drop in BLEU-4 scores, which plummet by over 22 points on the Spider datasets and 20.77 points on Bird dev.
Similarly, removing \textbf{Supervised Fine-Tuning} (\texttt{w/o SFT}) results in a non-functional model, with all metrics dropping to near zero. 
This confirms that SFT acts as the necessary adaptation step, bridging the general domain knowledge acquired during CPT to the specific format and requirements of the comment generation task.


\begin{tcolorbox}
\textbf{[RQ2]: The results confirm each stage is essential. CPT provides foundational knowledge and SFT performs task adaptation, together creating a competent baseline. DPO is then the critical step, aligning the model with nuanced human preferences.}
\end{tcolorbox}

\begin{table*}[!ht]
\centering
\caption{Refined error analysis statistics for SQL-Commenter, based on 37 error samples from Spider and 31 from Bird.}
\label{tab:error_analysis_refined}
\begin{tabular}{llcc}
\toprule
\textbf{Main Category} & \textbf{Sub-category} & \textbf{Spider (N=37)} & \textbf{Bird (N=31)} \\
\midrule
\multirow{4}{*}{\parbox{3cm}{\textbf{A. Semantic Logic Errors}}} 
& A1. Complex Structure Misinterpretation & 7 & 9 \\
& A2. Join Type Confusion & 8 & 4 \\
& A3. Aggregation/Filtering Logic Error & 7 & 3 \\
\cmidrule(lr){2-4}
& \textit{Subtotal} & \textbf{22 (59.5\%)} & \textbf{16 (51.6\%)} \\
\midrule
\multirow{4}{*}{\parbox{3cm}{\textbf{B. Omission of Information}}} 
& B1. Missing Filtering Conditions & 5 & 6 \\
& B2. Missing Ordering/Limitation Logic & 4 & 4 \\
& B3. Missing Edge Case Handling & 2 & 2 \\
\cmidrule(lr){2-4}
& \textit{Subtotal} & \textbf{11 (29.7\%)} & \textbf{12 (38.7\%)} \\
\midrule
\multirow{3}{*}{\parbox{3cm}{\textbf{C. Factual Inaccuracy}}} 
& C1. Entity Name Hallucination & 2 & 1 \\
& C2. Business Logic Hallucination & 2 & 2 \\
\cmidrule(lr){2-4}
& \textit{Subtotal} & \textbf{4 (10.8\%)} & \textbf{3 (9.7\%)} \\
\midrule
\textbf{Total} & & \textbf{37 (100\%)} & \textbf{31 (100\%)} \\
\bottomrule
\end{tabular}
\end{table*}

\subsubsection{RQ3: Human Evaluation}


To assess practical quality for RQ3, we conducted a human evaluation comparing SQL-Commenter against two top baselines: Qwen3-14B and DeepSeek-R1-Distill-Qwen-32B. Six SQL experts anonymously rated comments on \textbf{Correctness}, \textbf{Completeness}, and \textbf{Naturalness} (4-point scale) for 200 Spider and 100 Bird samples, which were selected via stratified sampling to ensure a representative comparison that preserves the test sets' difficulty distribution. As shown in Table~\ref{tab:human_eval}, SQL-Commenter significantly outperformed the baselines. On Spider, its Correctness positive rating (score $\geq$3) was \textbf{87.5\%} versus 67.5\%-69.5\% for baselines. The advantage held on the more complex Bird dataset, where our model scored \textbf{76.0\%} against their 53.0\%-58.0\%. Qualitative analysis showed that baselines often make logical omissions or factual errors on complex queries. For example, they might ignore a "per-region" clause in a window function or misinterpret a \texttt{LEFT JOIN ... IS NULL} for finding non-matching records as one that lists all records. In contrast, our model's high Completeness scores (\textbf{83.0\%} on Spider, \textbf{70.0\%} on Bird) highlight its superior ability.


\begin{tcolorbox}
\textbf{[RQ3]: Human evaluation confirms that SQL-Commenter's comments are more correct, complete, and natural than top-tier baselines. Its advantage is particularly pronounced in its reliability when handling complex analytical queries, effectively avoiding logical omissions and factual errors.}
\end{tcolorbox}

\subsubsection{RQ4: Error Analysis}
\label{sec:error_analysis}


To answer \textbf{RQ4}, we performed an error analysis to identify model limitations. We defined an error sample as any comment rated 1 or 2 (out of 4) by human experts on either Correctness or Completeness. This process identified 37 unique errors from the Spider dataset and 31 from Bird, which we categorized into three main types (Table~\ref{tab:error_analysis_refined}).

\paragraph{Insights from Error Analysis.}
The error distribution in Table~\ref{tab:error_analysis_refined} provides critical insights into the efficacy of our training methodology. The success of our CPT and DPO pipeline is most powerfully demonstrated by the scarcity of \textbf{C. Factual Inaccuracy} errors (approx. 10\%). These are critical failures where the model might hallucinate a non-existent column name (C1) or fabricate business logic not present in the SQL (C2). The low incidence of this category reflects our DPO's effectiveness in enforcing trustworthiness. 



Our approach also effectively reduced B. Omission of Information errors (29.7\% on Spider, 38.7\% on Bird). Remaining instances are typically context-dependent, where the model omits secondary clauses like `ORDER BY` or `LIMIT` clause (B2) in long queries, suggesting a limited attention span over extended contexts.

By effectively suppressing the most critical factual errors and narrowing the scope of omissions, the analysis brings the core remaining challenge into sharp focus: \textbf{A. Semantic Logic Errors}. This category now stands as the dominant failure mode, constituting the majority of errors on both Spider (59.5\%) and Bird (51.6\%). These errors stem from the model fundamentally misunderstanding the SQL's operational logic. For instance, the model may fail to explain the cascading filtering effect of a `LEFT JOIN` followed by an `INNER JOIN` (A2), misinterpret a correlated subquery as a static, independent calculation (A1), or confuse the nuanced logic of a window function like `RANK()` with a simple `GROUP BY` aggregation (A3). These specific mistakes point to a fundamental limitation in compositional reasoning. This suggests that while our training pipeline excels at enforcing preferences for factuality and completeness, resolving these deep logical misunderstandings pushes the boundaries of the 7B model's inherent capabilities. This points towards future work in leveraging larger base models or exploring advanced alignment techniques like GRPO, which are specifically designed to enhance multi-step reasoning.

\vspace{1em} 

\begin{tcolorbox}
\textbf{[RQ4]: The primary failure mode is misinterpreting complex semantic logic. Our training successfully minimized factual errors and omissions, but struggles with compositional reasoning, a core limitation tied to the 7B model's scale. Future work should target larger models and advanced alignment methods (e.g., GRPO) to improve reasoning capabilities.}
\end{tcolorbox}

\section{Related Work}


\subsection{Text-to-SQL Generation}

Early Text-to-SQL (NL2SQL) approaches used semantic parsing with symbolic grammars~\cite{wang2020rat, zhong2017seq2sql}, but faced scalability limits~\cite{li2014constructing}. They were largely superseded by end-to-end neural methods using Pre-trained Language Models (PLMs)\cite{guo2019towards, chen2020bridging, scholak2021picard, rubin2021smbop}. Research to enhance these PLMs has focused on relation-aware encoding\cite{wang2020rat, rubin2021smbop}, grammar-based decoding~\cite{yin2017syntactic, scholak2021picard}, schema linking~\cite{stoica2021re}, SQL sketching~\cite{cao2021lgesql}, and execution-guided generation~\cite{zhong2020semantic}.
More recently, Large Language Models (LLMs) have shifted the paradigm to prompt-driven generation. Initial studies evaluated the zero-shot capabilities of models like Codex~\cite{chen2021evaluating} and explored robustness enhancements via example selection~\cite{rajkumar2022evaluating, zhuo2023robustness}. Liu et al.\cite{liu2023comprehensive} found that ChatGPT generalizes well to unseen databases, though it still lags behind fine-tuned PLMs. Efforts to improve prompt quality include question classification\cite{yu2023grappa}, skeleton retrieval~\cite{cao2021lgesql}, and demonstration synthesis~\cite{huang2023generationsql}. Other advanced methods guide step-by-step generation with sub-task examples (DIN-SQL~\cite{pourreza2023din}) or combine PLM-generated sketches with value alignment for efficient generation (ZeroNL2SQL~\cite{fan2024combining}).

\subsection{SQL-to-Text Generation}


In the field of SQL-to-Text generation, graph-based methods have been explored to model SQL's structural properties. Graph Neural Networks~\cite{yu2023money, 
zhang2025topology, zhangrestricted} have also been applied to SQL-related tasks. Xu et al.~\cite{xu2018sql2text} proposed a graph-to-sequence model that leverages the structural information of SQL queries to generate natural language descriptions. To better capture SQL's complex syntax, Zhang et al.~\cite{zhang2025hesqlnet} introduced HeSQLNet, which uses a heterogeneous graph to explicitly distinguish between different types of structural relationships.
More recently, Zhang et al.~\cite{zhang2025semanticcaptioning} introduced the task of semantic captioning for SQL and constructed a benchmark dataset using iterative prompting with GPT-4o. They also proposed AST-ICL, a graph-aware few-shot demonstration selection method that improves generation performance, especially for small-scale models. In parallel, Zhang et al.~\cite{zhang2024comprehensive} incorporated SQL-to-Text as one of the evaluation sub-tasks in their Text-to-SQL benchmarking framework. Their experiments showed that general-purpose LLMs outperform coding-specific models when generating natural language from SQL.

\section{Threats to Validity}

\subsection{Internal Validity}
A threat to internal validity is the model's occasional tendency to generate repetitive sentences. We mitigate this by implementing a rule-based post-processing step that truncates the output upon detecting such repetitions. While this heuristic effectively handles straightforward cases, it might not address more subtle forms of redundancy. Future work will aim to improve the model's core generation process to prevent repetition at the source, thus reducing the reliance on post-processing rules.

\subsection{External Validity}
External validity is threatened by the scope of our training data. Our dataset was meticulously constructed using LLMs and refined by human experts, including database administrators and data analysts with over five years of experience. Although this dataset covers a wide range of complexities through the Spider and Bird benchmarks, it may not encompass all real-world analytical queries, such as the specific features of Oracle or SQL Server. Consequently, the model's performance on highly specialized or proprietary SQL variants remains a potential limitation.

\section{Conclusion}
In this paper, we propose SQL-Commenter, an advanced SQL code comment generation approach based on LLMs. Unlike prior work, SQL-Commenter employs a three-stage post-training process: CPT, SFT, and DPO. In addition, we constructed and publicly released comprehensive datasets for CPT, SFT, and DPO, significantly improving the quality of generated comments. Our approach outperforms state-of-the-art methods across automatic metrics (BLEU-4, METEOR, and ROUGE-L) on both the Spider and Bird benchmarks. Human evaluation further demonstrates that SQL-Commenter can provide correct, comprehensive, and natural comments.

\section*{ACKNOWLEDGMENTS}
This work was supported by the Alliance of International Science Organizations Collaborative Research Program (No.ANSO-CR-KP-2022-03).

\bibliographystyle{ACM-Reference-Format}
\bibliography{sample-base}










\end{document}